arXiv:1709.10278v2 [cs.GT] 9 Oct 2017

# Shapley Facility Location Games

Omer Ben-Porat and Moshe Tennenholtz

Technion - Israel Institute of Technology, Haifa, Israel
omerbp@campus.technion.ac.il
moshet@ie.technion.ac.il

**Abstract.** Facility location games have been a topic of major interest in economics, operations research and computer science, starting from the seminal work by Hotelling. Spatial facility location models have successfully predicted the outcome of competition in a variety of scenarios. In a typical facility location game, users/customers/voters are mapped to a metric space representing their preferences, and each player picks a point (facility) in that space. In most facility location games considered in the literature, users are assumed to act deterministically: given the facilities chosen by the players, users are attracted to their nearest facility. This paper introduces facility location games with probabilistic attraction, dubbed *Shapley facility location games*, due to a surprising connection to the Shapley value. The specific attraction function we adopt in this model is aligned with the recent findings of the behavioral economics literature on choice prediction. Given this model, our first main result is that Shapley facility location games are potential games; hence, they possess pure Nash equilibrium. Moreover, the latter is true for any compact user space, any user distribution over that space, and any number of players. Note that this is in sharp contrast to Hotelling facility location games. In our second main result we show that under the assumption that players can compute an approximate best response, approximate equilibrium profiles can be learned efficiently by the players via dynamics. Our third main result is a bound on the Price of Anarchy of this class of games, as well as showing the bound is tight. Ultimately, we show that player payoffs coincide with their Shapley value in a coalition game, where coalition gains are the social welfare of the users.

## 1 Introduction

In his seminal work [14], Hotelling considers a competition between two ice-cream vendors, who sell ice-cream to sunbathers on the beach, and wish to maximize their payoffs. The vendors sell the same type of product, and charge the same price. Sunbathers are distributed uniformly along the beach and every sunbather walks to his/her nearest ice-cream vendor to buy an ice-cream. As indicated by Hotelling, the vendors will strategically locate their ice-cream carts in the middle of the beach, back to back, as this is the only Nash equilibrium of this game.

Following that seminal work, facility location games have been a topic of major interest in economics, operations research and computer science. Spatial facility location models have successfully predicted the outcome of competition in a variety of scenarios. In a typical facility location game, users/customers/voters are mapped to a metric space representing their preferences, and each player picks a point (facility) in that space. Thereupon, each player is awarded one monetary unit for each user attracted to her facility. Even a toy example like the one above supports powerful real-world phenomena.

In most facility location games considered in the literature, users are assumed to act deterministically: given the facilities chosen by the players, users are attracted to their nearest facility. Indeed, such rational behavior of users is justified in many situations. However, far too little attention has been paid to models where users are not deterministic, and are not simply attracted to their nearest facility. Irrational decision making is ubiquitous, as demonstrated by the celebrated work of Kahneman and Tversky [16]. In this context, analyzing probabilistic user attraction introduces new theoretical challenges to overcome, as well as practical implications.

This paper focuses on facility location games with probabilistic attraction. Our proposed attraction function is aligned with the "Satisficing" principle in decision making [30], and the model of selection based on small samples [11,3]. The specific attraction function we adopt can be found in the recent experimental economics benchmark presented in [10], and its usefulness in choice prediction is discussed in [21].

We first formally present the above modeling process to determine the attraction probabilities. Using this attraction, we define the class of facility location games considered in this paper, termed *Shapley facility location games*, due to a surprising connection to the Shapley value [28]. The difference between our model and Hotelling's is analyzed using the toy example above; in particular, we show that when both players choose the middle of the segment, this is no longer an equilibrium profile; indeed, facilities will be selected and located in different locations.

We then show that Shapley facility location games are potential games [19]; hence, they possess pure Nash equilibrium. Moreover, the latter is true for any compact user space, any user distribution over that space, and any number of players. Note that this is in sharp contrast to Hotelling facility location games, where pure Nash equiibrium does not always exist (see, e.g., [27,26,9]).

An interesting question is whether strategic interaction among the players will converge to an approximate Nash equilibrium (see, e.g., [7,2]). The dynamics of Hotelling facility location games refer to intractable problems, and is rarely analyzed. We show that under the assumption that players can compute an approximate best response, approximate equilibrium profiles can be learned efficiently by the players via dynamics in any Shapley facility location game. This result holds for any user space (including an infinite one). We also bound the Price of Anarchy [17,24] of this class of games, and show the bound is tight.

Ultimately, the connection to the Shapley value is provided, as we bind (non-cooperative) facility location games with our selection of probabilistic attraction to cooperative game theory. We show that player payoffs coincide with their Shapley values in a coalition game, where coalition gains are the social welfare of the users.

### 1.1 Related Work

For a recent survey of Hotelling games the reader is referred to [5]. In the same spirit, Voronoi games (see, e.g., [1,6,12,1]) look at the game theoretic aspects of facility location with potentially multiple facilities for each player in general (euclidean) spaces.

The above work does not refer to probabilistic selection among facilities, an essential aspect needed in order to deal with realistic commerce and marketing setups. An exception that does adopt some form of probabilistic selection is the model of [13]. We will discuss how [13] can be seen as a special case of our model in Section 6.

Probabilistic choice among products [18] is widely explored, and choice prediction [30,11,3] is studied extensively. In this line of work, the authors wish to predict how a subject will make his/her choice among products. In our paper the way users react given a set of products is adopted from that literature, and embedded in the context of facility location games.

A different line of research in the algorithmic game theory literature is the study of facility location in the context of approximate mechanism design [20]. That literature deals with the case where only one entity dictates the place of a facility (or several facilities), while user preferences are their private information and are strategically reported, see e.g. [25,22]. In that context the players are the users, while our work extends facility location games where the players are the facilities' owners.

## 2 Model

Before we present our model formally, we briefly describe a general facility location game, and elaborate on the component we revisit.

Typically in a facility location game, users are distributed in a space $\mathcal{U}$, where every point $u \in \mathcal{U}$ models a user, be it by his[1] physical location, his preferences towards a product, or his political point of view. The space $\mathcal{U}$ plays one more role: every point in $\mathcal{U}$ is also a potential location for a *facility*, which is a physical location of a store, properties of a product, or political agenda. There are $n$ players, where each player is to locate her facility in $\mathcal{U}$. Namely, a strategy of player $i$ is a location $x_i \in \mathcal{U}$. A strategy profile is a vector describing where each player located her facility, $\boldsymbol{x} = (x_1, \dots, x_n) \in \mathcal{U}^n$.

Each user $u \in \mathcal{U}$ has a similarity function $\sigma_u : \mathcal{U} :\to [0, 1]$, where $\sigma_u(t)$ quantifies the extent to which $u \in \mathcal{U}$ is satisfied with a facility located in $t \in U$.[2] Given a strategy profile $\boldsymbol{x}$, users are *attracted* to the facilities of the players according to *some* attraction function, which receives the *similarity vector* $\sigma_u(\boldsymbol{x}) = (\sigma_u(x_1), \dots, \sigma_u(x_n))$ as input.[3] Players are strategic, namely they locate their facilities with the aim of attracting as many users as possible.

The component we revisit in this paper is the attraction function. Following the behavioral economics literature, users do not just select the facility they are most satisfied with (e.g. are not simple expected utility maximizers [16]). In this work we focus on the analysis of facility location games with a user attraction function that is popular in behavioral science as described in the introduction, thereby incorporating the human aspect in our model. Indeed, it has been shown that this modelling is an extremely effective ingredient in the context of choice prediction [10,21].

Given the locations selected by the players, the process of deciding which facility to select, if any, is modeled as follows: every user samples a satisfaction threshold from the uniform distribution[4], and then chooses a facility with satisfaction level above that threshold, if such a facility exists. If several facilities meet his criterion, he flips an unbiased coin to remain with one facility.

Surprisingly, as we show in Section 5, the aforementioned simple and intuitive selection process leads to a standard solution concept in cooperative game theory. More precisely, the probability of $u$ to select facility $x_i$ coincides with the Shapley value of player $i$ in a cooperative game where the value assigned to each coalition is the maximal similarity level of $u$ with the facilities of that coalition. For that reason, we term it the *Shapley attraction function.* A formal definition of the Shapley attraction function is as follows.

**Definition 1.** *For a strategy profile $\boldsymbol{x}$ and a user $u$, let $\left(\sigma_u^1(\boldsymbol{x}), \sigma_u^2(\boldsymbol{x}), \dots \sigma_u^n(\boldsymbol{x})\right)$ denote the result of ordering the similarity vector $\sigma_u(\boldsymbol{x})$ in ascending order, and let $\rho_i = \rho_i(u, \boldsymbol{x})$ be an index such that $\sigma_u(x_i) = \sigma_u^{\rho_i}(\boldsymbol{x})$. Under the Shapley attraction function, $u$ is attracted to each*

[1] For ease of exposition, third-person singular pronouns are "she" for a player and "he" for a user.
[2] Commonly in facility location models, distances are used to determine the attraction. However, for ease of presentation of the model and the analysis, we employ proximity; clearly, both notions are equivalent.
[3] In Hotelling games, for instance, each user $u$ selects a player uniformly from $\{i : \sigma_u(x_i) \geq \max_j \sigma_u(x_j)\}$.
[4] Our results hold for any distribution, as well as in case the distribution is different for each user.

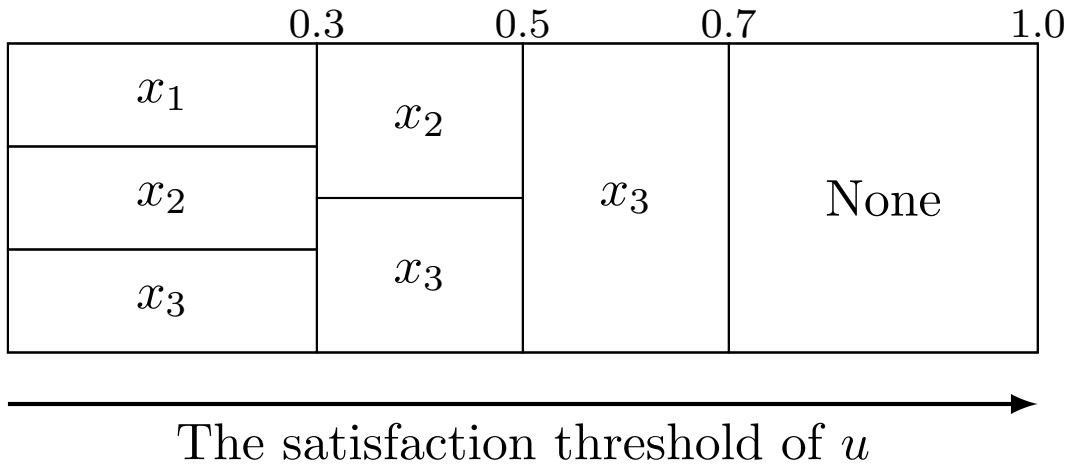

**Fig. 1.** Consider a user $u$ and a strategy profile $\boldsymbol{x} = (x_1, x_2, x_3)$ such that the similarity vector $\sigma_u(\boldsymbol{x}) = (\sigma_u(x_1), \sigma_u(x_2), \sigma_u(x_3)) = (0.3, 0.5, 0.7)$. Hence, $\left(\sigma_u^1(\boldsymbol{x}), \sigma_u^2(\boldsymbol{x}), \sigma_u^3(\boldsymbol{x})\right) = (0.3, 0.5, 0.7)$ as well. User $u$ samples his satisfaction threshold $Y$ (as mentioned, uniformly distributed random variable). If $Y \leq 0.3$, then all the facilities satisfy him, so he chooses one uniformly. If $0.3 < Y \leq 0.5$, only $x_2, x_3$ satisfy him, and so he flips a coin to choose one of them. If $0.5 < Y \leq 0.7$, the only satisfying facility is $x_3$, and if $Y > 0.7$ he will not select any facility. It follows that $u$ will select $x_1$ with probability $\mu_1(u, \boldsymbol{x}) = \frac{\sigma_i^1}{3} = \frac{1}{10}$, $x_2$ with probability $\mu_2(u, \boldsymbol{x}) = \frac{\sigma_i^1}{3} + \frac{\sigma_i^2 - \sigma_i^1}{2} = \frac{2}{10}$, and $x_3$ with probability $\mu_3(u, \boldsymbol{x}) = \frac{\sigma_i^1}{3} + \frac{\sigma_i^2 - \sigma_i^1}{2} + \frac{\sigma_i^3 - \sigma_i^2}{1} = \frac{4}{10}$. With probability 0.3 he will select none of the facilities.

*player $i$ with probability*

$$\mu_i(u, \boldsymbol{x}) \triangleq \Pr\left(u \textit{ is attracted to } i \textit{ under } \boldsymbol{x}\right) = \sum_{j=1}^{\rho_i} \frac{\sigma_u^j(\boldsymbol{x}) - \sigma_u^{j-1}(\boldsymbol{x})}{n - j + 1}, \tag{1}$$

*where* $\sigma_i^0(\boldsymbol{x}) = 0$.

See Figure 1 for illustration. We are now ready to formally present the model. A Shapley facility location game is composed of the following:

1. A compact set of users $\mathcal{U}$, and a density function $f$ with mass 1 over $\mathcal{U}$.
2. A similarity function $\sigma : \mathcal{U} \times \mathcal{U} :\to [0, 1]$, such that $\sigma_u(t) \triangleq \sigma(u, t)$ for all $t \in \mathcal{U}$.
3. A set of players, $[n] = \{1, \ldots, n\}$. The strategy set of each player $i$ is a location (facility) in $\mathcal{U}$. The strategy of player $i$ is denoted by $x_i \in \mathcal{U}$, and a strategy profile by $\boldsymbol{x} = (x_1, \ldots, x_n) \in \mathcal{U}^n$.
4. Users are attracted to player facilities according to the Shapley attraction function. That is, the probability that $u$ will be attracted to facility $x_i$ of player $i$ under $\boldsymbol{x}$ is $\mu_i(u, \boldsymbol{x})$ given in Equation (1).
5. The payoff of player $i$ under the strategy profile $\boldsymbol{x}$ is the proportion of users attracted to her chosen location, i.e.
$$\pi_i(\boldsymbol{x}) = \int_{\mathcal{U}} f(u)\mu_i(u, \boldsymbol{x})du. \tag{2}$$

Throughout the paper, both $\sigma_u(\cdot)$ and $\sigma(u, \cdot)$ are used interchangeably. We restrict the scope of this work to similarity functions that are Riemann integrable, for instance continuous functions or simple functions (a finite linear combination of indicator functions). In euclidean spaces, natural similarity functions are monotonically non-increasing in the distance. Note, however, that a similarity function need not be monotone.

We say that a strategy profile $\boldsymbol{x} = (x_1, \ldots, x_n) \in \mathcal{U}^n$ is a *pure Nash equilibrium* if for any player $i \in [n]$ and any strategy $x_i' \in \mathcal{U}$ it holds that $\pi_i(x_i, \boldsymbol{x}_{-i}) \geq \pi_i(x_i', \boldsymbol{x}_{-i})$, where $\boldsymbol{x}_{-i}$ denotes the vector $\boldsymbol{x}$ of all strategies, but with the $i$-th component deleted.

## 3 An Illustrative Example

In this section we illustrate Shapley facility location games by considering a game instance, thereby demonstrating the elements of the model. We employ the very restricted two-player, uniform distribution on a segment setting considered in [14]. We stress that this section serves as a demonstration only, and our results in the upcoming section apply to the model described above in its full generality.

We focus on a game $\mathcal{G}$ induced by the space of users $\mathcal{U} = [0,1]$, uniform probability distribution $f(u) = \mathbb{1}_{0 \leq u \leq 1}$, two players, and a symmetric similarity function

$$\forall u, t \in [0,1] : \sigma(u,t) = 1 - |u - t|.$$

Note that $\sigma(u,t)$ is merely one minus the absolute distance between $u \in [0,1]$ and a potentially occupied location $t \in [0,1]$.

Let $(x_1, x_2)$ be a strategy profile such that $x_1 \leq x_2$. Observe that

$$\mu_1\left(u, (x_1, x_2)\right) = \begin{cases} \frac{\sigma(u,x_2)}{2} + \sigma(u,x_1) - \sigma(u,x_2) & u < \frac{x_1+x_2}{2} \\ \frac{\sigma(u,x_1)}{2} & u \geq \frac{x_1+x_2}{2} \end{cases}.$$

See Figure 2 for visualization of the above. The payoff of player 1 is given by

$$\pi_1(x_1,x_2) = \int_0^1 \mu_1\left(u,(x_1,x_2)\right) du = \int\limits_0^{\frac{x_1+x_2}{2}} \left(\sigma(u,x_1) - \frac{\sigma(u,x_2)}{2}\right) du + \int\limits_{\frac{x_1+x_2}{2}}^{1} \sigma(u,x_1) du.$$

The construction of player 2's payoff is similar. Using elementary calculations, one can find the pure Nash equilibria of $\mathcal{G}$.

**Proposition 1.** *The strategy profile* $\left(\frac{3}{8}, \frac{5}{8}\right)$ *is the unique pure Nash equilibrium of* $\mathcal{G}$*, up to renaming the players.*

The proof of Proposition 1 is in the appendix. Indeed, in contrast to [14], under equilibrium profile players choose different locations. We leave the complete analysis of this setting (i.e. more players, higher dimensional space) for future work.

## 4 Analysis

We now examine the properties of Shapley facility location games. We begin with showing that every Shapley facility location game possesses a pure Nash equilibrium. Afterwards, we show that if mild assumptions are satisfied, learning dynamics will efficiently converge to an approximate equilibrium. This is despite of the infinite strategy space of the players. Finally, the price of anarchy is analyzed.

### 4.1 Pure Nash Equilibrium

In this subsection we show that Shapley facility location games possess pure Nash equilibrium. A non-cooperative game is called a *potential game* [19] if there exists a function $\Phi : \mathcal{U}^n \to \mathbb{R}$ such that for every strategy profile $\boldsymbol{x} = (x_1, \ldots, x_n) \in \mathcal{U}^n$ and every $i \in [n]$, whenever player $i$ switches from $x_i$ to a strategy $x_i' \in \mathcal{U}$, the change in her payoff function equals the change in the potential function, i.e.

$$\Phi(x_i', \boldsymbol{x}_{-i}) - \Phi(x_i, \boldsymbol{x}_{-i}) = \pi_i(x_i', \boldsymbol{x}_{-i}) - \pi_i(x_i, \boldsymbol{x}_{-i}).$$

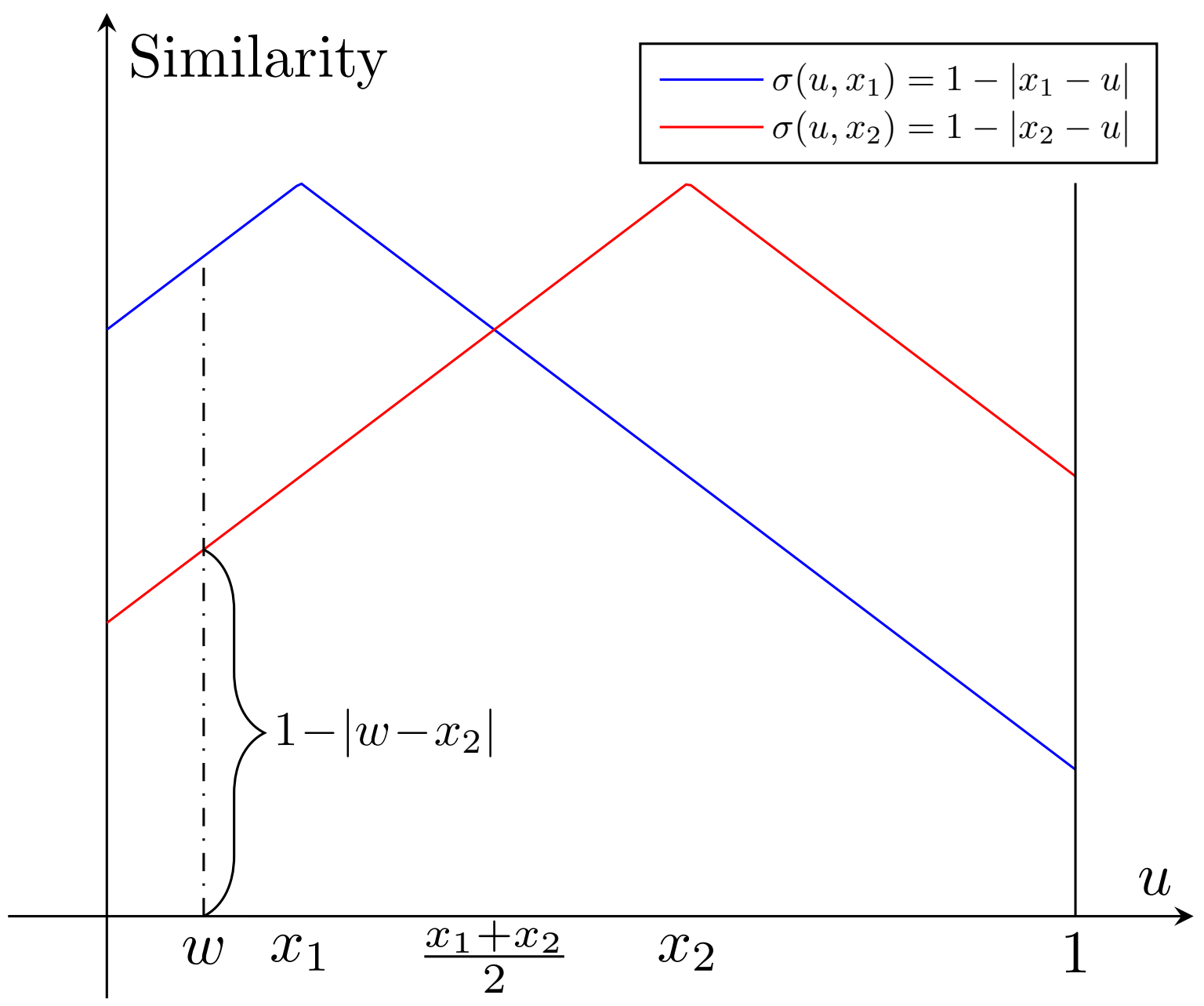

**Fig. 2.** User similarity with respect to the strategy profile $(x_1, x_2)$. The blue line is the similarity function of $x_1$ with every user, and the red line is that of $x_2$. Every user $w \leq \frac{x_1+x_2}{2}$ selects $x_1$ with probability $\mu_1(w, (x_1, x_2)) = \frac{1-|x_2-w|}{2} + (|x_2 - w| - |x_1 - w|)$ and $x_2$ with probability $\mu_2(w, (x_1, x_2)) = \frac{1-|x_2-w|}{2}$. Similarly, every user $v \geq \frac{x_1+x_2}{2}$ selects $x_1$ with probability $\mu_1(v, (x_1, x_2)) = \frac{1-|x_1-v|}{2}$ and $x_2$ with probability $\mu_2(v, (x_1, x_2)) = \frac{1-|x_1-v|}{2} + (|x_1 - v| - |x_2 - v|)$.

**Theorem 1.** *Shapley facility location games are potential games.*

*Proof.* Fix a player $i$. Given a strategy profile $\boldsymbol{x}$, define:

$$c_u(y; \boldsymbol{x}) = |\{i \in [n] : y \leq \sigma_u(x_i)\}|.$$

The latter represents the number of players that attract the infinitesimal user $u$ under the locations of the players defined by the profile $\boldsymbol{x}$, in case he sampled the satisfaction level $y$. Consequently, the payoff of player $i$, formerly defined in Equation (2), can be reformulated as

$$\pi_i(\boldsymbol{x}) = \int_{\mathcal{U}} f(u) \int_0^{\sigma_u(x_i)} \frac{1}{c_u(y; \boldsymbol{x})} dydu. \tag{3}$$

Next, we show that

$$\Phi(\boldsymbol{x}) = \int_{\mathcal{U}} f(u) \int_{y=0}^{1} \sum_{i=1}^{c_u(y;\boldsymbol{x})} \frac{1}{i} dydu$$

is a potential function of the game. We temporarily focus on a user $u$. For any strategy profile $\boldsymbol{x}$ and user $u$, it holds that

$$c_u(y; \boldsymbol{x}) = \begin{cases} c_u(y; \boldsymbol{x}_{-i}) & y > \sigma_u(x_i) \\ c_u(y; \boldsymbol{x}_{-i}) + 1 & y \leq \sigma_u(x_i) \end{cases}.$$

Therefore, we have

$$
\begin{aligned}
&\int_0^{\sigma_u(x_i)} \frac{dy}{c_u(y;\boldsymbol{x})} + \int_0^1 \sum_{j=1}^{c_u(y;\boldsymbol{x}_{-i})} \frac{1}{j} dy \\
&= \int_0^{\sigma_u(x_i)} \frac{dy}{c_u(y;\boldsymbol{x}_{-i})+1} + \int_0^{\sigma_u(x_i)} \sum_{j=1}^{c_u(y;\boldsymbol{x}_{-i})} \frac{1}{j} dy + \int_{\sigma_u(x_i)}^1 \sum_{j=1}^{c_u(y;\boldsymbol{x}_{-i})} \frac{1}{j} dy \\
&= \int_0^{\sigma_u(x_i)} \sum_{j=1}^{c_u(y;\boldsymbol{x}_{-i})+1} \frac{1}{j} dy + \int_{\sigma_u(x_i)}^1 \sum_{j=1}^{c_u(y;\boldsymbol{x}_{-i})} \frac{1}{j} dy \\
&= \int_0^1 \sum_{j=1}^{c_u(y;\boldsymbol{x})} \frac{1}{j} dy. \qquad (4)
\end{aligned}
$$

We are now ready for the final argument. Fix two profiles, $(x_i, \boldsymbol{x}_{-i}), (x'_i, \boldsymbol{x}_{-i})$. It follows that

$$
\begin{aligned}
&\pi_i(x_i, \boldsymbol{x}_{-i}) - \pi_i(x'_i, \boldsymbol{x}_{-i}) = \\
&\int_{\mathcal{U}} f(u) \int_0^{\sigma_u(x_i)} \frac{1}{c_u(y; x_i, \boldsymbol{x}_{-i})} dydu - \int_{\mathcal{U}} f(u) \int_0^{\sigma_u(x'_i)} \frac{1}{c_u(y; x'_i, \boldsymbol{x}_{-i})} dydu = \\
&\int_{\mathcal{U}} f(u) \int_0^{\sigma_u(x_i)} \frac{1}{c_u(y; x_i, \boldsymbol{x}_{-i})+1} dydu - \int_{\mathcal{U}} f(u) \int_0^{\sigma_u(x'_i)} \frac{1}{c_u(y; x'_i, \boldsymbol{x}_{-i})+1} dydu. \qquad (5)
\end{aligned}
$$

By adding and removing $\int_{\mathcal{U}} f(u) \int_0^1 \sum_{j=1}^{c_u(y;\boldsymbol{x}_{-i})} \frac{1}{j} dydu$ to Equation (5), similar to what we showed in Equation (4), we obtain

$$
\begin{aligned}
(5) &= \int_{\mathcal{U}} f(u) \int_0^1 \sum_{j=1}^{c_u(y;\boldsymbol{x})} \frac{1}{j} dydu - \int_{\mathcal{U}} f(u) \int_0^1 \sum_{j=1}^{c_u(y;x'_i,\boldsymbol{x}_{-i})} \frac{1}{j} dydu \\
&= \Phi(\boldsymbol{x}) - \Phi(x'_i, \boldsymbol{x}_{-i}).
\end{aligned}
$$

□

Since $\mathcal{U}$ is a compact set and the payoff functions are continuous with respect to the strategy space, a direct result from Theorem 1 and [19, Lemma 4.3] is the following.

**Corollary 1.** *Every Shapley facility location game possesses a pure Nash equilibrium.*

### 4.2 Convergence to Approximate Equilibrium

In this subsection we examine learning dynamics of Shapley facility location games. The solution concept we are after is (multiplicative) approximate pure Nash equilibrium. In [7], the authors examined convergence of dynamics in symmetric (finite) congestion games. However, in Shapley facility location games the user space may be infinite; hence modifications are needed.

We begin with a few definitions. We say that a strategy profile $\boldsymbol{x}$ is an $\epsilon$-pure Nash equilibrium ($\epsilon$-PNE) for $\epsilon > 0$ if

$$
\forall i \in [n], \forall x'_i \in \mathcal{U}: \quad \pi_i(x'_i, \boldsymbol{x}_{-i}) \leq (1+\epsilon)\pi_i(\boldsymbol{x}).
$$

Notice that if $\boldsymbol{x}$ is an $\epsilon$-PNE, then any player cannot improve her payoff by a factor of more than $(1+\epsilon)$ of what she gets under $\boldsymbol{x}$ by unilaterally deviating to another location.

In the upcoming analysis, we assume players can efficiently compute $\epsilon$-best response, if such exists. Indeed, this assumption holds for several plausible scenarios, such as concave payoff functions or discretization of Lipschitz user distribution.

The dynamics we consider are the following:

**Best-response dynamics**:

- Until reaching $\epsilon$-PNE:
  - Pick an arbitrary player with a $(1+\epsilon)$ profitable deviation, and move her to her deviating strategy.

It turns out that any such strategic interaction among the players will converge to an $\epsilon$-PNE after efficient number of iterations.

**Theorem 2.** *Let $\epsilon \in (0,1)$. In a Shapley facility location game with $n$ players and an initial strategy $\boldsymbol{x}_0$, after $\mathcal{O}\left(\frac{n \log n}{\epsilon} \log \frac{\Phi_{\max}}{\Phi(\boldsymbol{x}_0)}\right)$ any best-response dynamics converges to $\epsilon$-PNE.*

Before we turn to prove Theorem 2, we prove two supporting lemmas.

**Lemma 1.** *For every profile $\boldsymbol{x}$ it holds that*

$$\sum_{i=1}^{n} \pi_i(\boldsymbol{x}) \geq \frac{\Phi(\boldsymbol{x})}{\ln(n)+1}.$$

*Proof.* Fix a strategy profile $\boldsymbol{x}$. Observe that

$$\begin{aligned}
\sum_{i=1}^{n} \pi_i(\boldsymbol{x}) &= \sum_{i=1}^{n} \int_{\mathcal{U}} f(u) \int_{0}^{\sigma_u(x_i)} \frac{1}{c_u(y;\boldsymbol{x})} dy du \\
&= \int_{\mathcal{U}} f(u) \sum_{i=1}^{n} \int_{0}^{\sigma_u(x_i)} \frac{1}{c_u(y;\boldsymbol{x})} dy du \\
&= \int_{\mathcal{U}} f(u) \int_{c_u(y;\boldsymbol{x}) \neq 0} 1 dy du.
\end{aligned}$$

Since $H_n = \sum_{i=1}^{n} \frac{1}{i} < \ln(n)+1$, we have:

$$\begin{aligned}
\int_{\mathcal{U}} f(u) \int_{c_u(y;\boldsymbol{x}) \neq 0} 1 dy du &\geq \frac{1}{\ln(n)+1} \int_{\mathcal{U}} f(u) \int_{c_u(y;\boldsymbol{x}) \neq 0} \sum_{i=1}^{n} \frac{1}{i} dy du \\
&\geq \frac{1}{\ln(n)+1} \int_{\mathcal{U}} f(u) \int_{0}^{1} \sum_{i=1}^{c_u(y;\boldsymbol{x})} \frac{1}{i} dy du \\
&= \frac{\Phi(\boldsymbol{x})}{\ln(n)+1}.
\end{aligned}$$

□

**Lemma 2.** *Denote by $i$ the index of the player chosen by the dynamics, and let $x_i'$ denote her deviation. It follows that:*

$$\forall j \in [n]: \quad \pi_i(x_i', \boldsymbol{x}_{-i}) - \pi_i(\boldsymbol{x}) \geq \frac{\epsilon}{4} \pi_j(\boldsymbol{x}).$$

*Proof.* In case $\pi_i(\boldsymbol{x}) \geq \frac{\pi_j(\boldsymbol{x})}{4}$, player $i$ has an $\epsilon$-profitable deviation; hence it holds that

$$\pi_i(x_i', \boldsymbol{x}_{-i}) - \pi_i(\boldsymbol{x}) \geq \epsilon \pi_i(\boldsymbol{x}) \geq \frac{\epsilon}{4} \pi_j(\boldsymbol{x}).$$

Otherwise, $\pi_i(\boldsymbol{x}) < \frac{\pi_j(\boldsymbol{x})}{4}$. Next, for every $u, y$ such that $y \leq \sigma_u(s_j)$ it holds that

$$\begin{aligned} c_u(y; s_j, \boldsymbol{x}_{-i}) &= 2c_u(y; s_j) + c_u(y; \boldsymbol{x}_{-i,j}) \leq 2c_u(y; s_j) + c_u(y; \boldsymbol{x}_{-i,j}) + c_u(y; x_i) \\ &= c_u(y; \boldsymbol{x}) + 1 \leq 2c_u(y; \boldsymbol{x}). \end{aligned}$$

Thus,

$$\begin{aligned} \pi_i(s_j, \boldsymbol{x}_{-i}) &= \int_{\mathcal{U}} f(u) \int_0^{\sigma_u(s_j)} \frac{1}{c_u(y; s_j, \boldsymbol{x}_{-i})} dy du \geq \int_{\mathcal{U}} f(u) \int_0^{\sigma_u(s_j)} \frac{1}{2c_u(y; \boldsymbol{x})} dy du \\ &= \frac{\pi_j(\boldsymbol{x})}{2}. \end{aligned}$$

Hence,

$$\pi_i(x_i', \boldsymbol{x}_{-i}) - \pi_i(\boldsymbol{x}) \geq \pi_i(s_j, \boldsymbol{x}_{-i}) - \pi_i(\boldsymbol{x}) \geq \frac{\pi_j(\boldsymbol{x})}{2} - \frac{\pi_j(\boldsymbol{x})}{4} \geq \frac{\epsilon}{4} \pi_j(\boldsymbol{x}).$$

□

We are now ready to prove Theorem 2.

*Proof (of Theorem 2).* In one iteration of the dynamics it holds that

$$\begin{aligned} \Phi(x_i', \boldsymbol{x}_{-i}) - \Phi(\boldsymbol{x}) &= \pi_i(x_i', \boldsymbol{x}_{-i}) - \pi_i(\boldsymbol{x}) \\ &\geq \frac{\epsilon}{4} \max_j \pi_j(\boldsymbol{x}) \\ &\geq \frac{\epsilon}{4n} \sum_{j=1}^{n} \pi_j(\boldsymbol{x}) \\ &\geq \frac{\epsilon}{4n\left(\ln(n) + 1\right)} \Phi(\boldsymbol{x}). \end{aligned}$$

Let $c = \frac{\epsilon}{4n(\ln(n)+1)} < 1$. Denote by $m$ the number of iterations until convergence. Observe that

$$\Phi_{\max} \geq \Phi(\boldsymbol{x}_m) \geq (1+c)^m \Phi(\boldsymbol{x}_0). \tag{6}$$

If $m$ does not satisfy Equation (6),

$$\Phi_{\max} < (1+c)^m \Phi(\boldsymbol{x}_0) \leq e^{m \cdot c} \Phi(\boldsymbol{x}_0) \Rightarrow m \geq \frac{4n\left(\ln(n) + 1\right)}{\epsilon} \ln\left(\frac{\Phi_{\max}}{\Phi(\boldsymbol{x}_0)}\right).$$

Therefore, an $\epsilon$-PNE is obtained after at most $\mathcal{O}\left(\frac{n \log n}{\epsilon} \log \frac{\Phi_{\max}}{\Phi(\boldsymbol{x}_0)}\right)$ iterations of any best response dynamics. □

### 4.3 Price of Anarchy

In this subsection we analyze the Price of Anarchy [17,24] of the discussed games, herein denoted $PoA$. The $PoA$ measures the inefficiency of a game in terms of social welfare, as a result of selfish behavior of the players. Specifically, it is the ratio between an optimal dictatorial scenario and the social welfare of the worst equilibrium. If $S$ is the set of all feasible profiles, and $E \subseteq S$ is the set of pure equilibrium profiles, then:

$$PoA = \frac{\max_{\boldsymbol{x} \in S} V(\boldsymbol{x})}{\min_{\boldsymbol{x} \in E} V(\boldsymbol{x})}.$$

The objective function of interest is the following:

$$V(\boldsymbol{x}) = \int_{\mathcal{U}} f(u) \max_i \sigma_u(x_i) du.$$

Note that $V$ represents the sum of payoffs of the players, as well as the weighted maximum similarity users attain from the facilities under $\boldsymbol{x}$.

**Theorem 3.** *The PoA of Shapley facility location games is at most* $\frac{2n-1}{n}$.

The proof is in the appendix. After bounding the $PoA$, our objective is to show that this bound is tight, by presenting a game instance that achieves this bound.

**Lemma 3.** *There exists a game instance with* $PoA = \frac{2n-1}{n}$.

*Proof.* Consider an $n$-player game over $\mathcal{U} = [0,2]^n$. Let $e_i$ denote the $i$'th vector of the canonical basis of $\mathbb{R}^n$, $\mathbf{0}$ be the zero vector in $\mathbb{R}^n$, and let $B_i = \{w \in \mathcal{U} : d(w, e_i) < \epsilon\}$ where $d(\cdot)$ is the euclidean distance and $\epsilon > 0$ is a small constant. Denote by $\alpha$ the volume of each such $B_i$. Consider the following density function:

$$f(u) = \begin{cases} \frac{1}{\alpha n} & \exists i : u \in B_i \\ 0 & Otherwise \end{cases}.$$

In addition, let the similarity function be

$$\forall u, w \in \mathcal{U} : \ \sigma_u(w) = \begin{cases} 1 & d(u,w) < \epsilon \text{ and } w \neq \mathbf{0} \\ \frac{n}{2n-1} & w = \mathbf{0} \\ 0 & Otherwise \end{cases}.$$

We now show that the strategy profile $\boldsymbol{x} = (\mathbf{0}, \mathbf{0}, \ldots, \mathbf{0})$ is in equilibrium. Consider player $i$'s payoff under $\boldsymbol{x}$, and a possible unilateral deviation of her to $e_i$:

$$\pi_i(\boldsymbol{x}) = \frac{n}{2n-1}\frac{1}{n} = \frac{1}{2n-1}, \ \pi_i(e_i, \boldsymbol{x}_{-i}) = \frac{1}{n}\left(\frac{n}{2n-1}\frac{1}{n} + 1 - \frac{n}{2n-1}\right) = \frac{1}{2n-1}.$$

Since strategies outside $\{\mathbf{0}, e_1, \ldots, e_n\}$ are strictly dominated, we obtain $\pi_i(\boldsymbol{x}) \geq \pi_i(w, \boldsymbol{x}_{-i})$ for all $w \in \mathcal{U}$. Observe that $V(\boldsymbol{x}) = \frac{n}{2n-1}$. The optimal social welfare is one, obtained when players select unique locations, e.g. player $i$ selects $e_i$. Therefore, $PoA = \frac{2n-1}{n}$. □

## 5 Relation to Shapley Value

Imagine a user being puzzled by the offers of the players. A novel way to decide which facility to select is to consider the players as being collaborative, and divide its share among all players, where each player gets a "fair" part. In this section, we show that the previously defined user reaction function coincides with a core solution concept in cooperative game theory, and can be characterized by a collection of desirable properties.

A cooperative game consists of two elements: a set of players and a *characteristic function*, which assigns a value to every coalition, i.e. every subset of players. The analysis of cooperative games focuses on predicting which coalitions will be formed, and how the payoff of a coalition should be distributed among its members. One core solution concept is the Shapley value [28].

**Definition 2 (Shapley value).** *Given a cooperative game with a set of players $[n]$ and a characteristic function $v : 2^{[n]} \to \mathbb{R}$ such that $v(\phi) = 0$, the Shapley value is a way to distribute the total gain among the players. According to the Shapley value, the amount that player $i$ gets in a coalition game $(v, [n])$ is:*

$$\phi_i(v) \triangleq \frac{1}{n!} \sum_{R \in \Pi([n])} \left[ v(P_i^R \cup \{i\}) - v(P_i^R) \right] \tag{7}$$

*where $\Pi([n])$ is the set of all permutations of $[n]$ and $P_i^R$ is the set of players which precede $i$ in the permutation $R$.*

The Shapley value is characterized by a collection of desirable properties:

- **Efficiency**: $\sum_{i=1}^{n} \phi_i(v) = v\left([n]\right)$, i.e. the total gain is distributed.
- **Null player**: If $\forall \mathcal{C} \subseteq [n]$ it holds that $v(\mathcal{C} \cup \{i\}) = v(\mathcal{C})$, then $\phi_i(v) = 0$.
- **Symmetry**: If $i$, $j$ are equivalent, namely $v(\mathcal{C} \cup \{i\}) = v(\mathcal{C} \cup \{j\})$ for all $\mathcal{C} \subseteq [n]$, then $\phi_i(v) = \phi_j(v)$.
- **Linearity**: If $v, w$ are two cooperative games and $\alpha$ is a real number, then $\phi_i(\alpha v + w) = \alpha\phi_i(v) + \phi_i(w)$.

For our purposes, we temporarily focus on a specific user $u$. The characteristic function $v_u(\mathcal{C}; \boldsymbol{x})$ is defined to be the maximum similarity of $u$ to one of the facilities chosen by the members of $u$ under $\boldsymbol{x}$. Formally:

$$v_u(\mathcal{C}; \boldsymbol{x}) = \max_{i \in \mathcal{C}} \sigma_u(x_i).$$

This modeling follows the logic of Hotelling games, where each user is attracted to his nearest facility. Therefore, each user $u$ initiates a cooperative game that consists of the players $[n]$, and $v_u(; \boldsymbol{x})$ as a characteristic function.

Denote the cooperative game defined over all users by $V$,

$$V(\mathcal{C}; \boldsymbol{x}) = \int_{\mathcal{U}} f(u) v_u(\mathcal{C}; \boldsymbol{x}) du.$$

We now bind the payoff of a player in the facility location model presented above and its Shapley value of the cooperative game $V$.

**Theorem 4.** *The payoff of player $i$ under any pure strategy profile $\boldsymbol{x}$ is her Shapley value in the cooperative game $([n], V(; \boldsymbol{x}))$. Namely,*

$$\pi_i(\boldsymbol{x}) = \phi_i\left(V(; \boldsymbol{x})\right).$$

*Proof.* Due to [28,8], the Shapley value is fully characterized by the properties above. Therefore, if we show that the Shapley attraction function satisfies these properties, the theorem will be proven. Fix a strategy profile $\boldsymbol{x}$ and a user $u$. We show that $\mu_i(u, \boldsymbol{x})$ is the Shapley value of the cooperative game $v_u(;\boldsymbol{x})$:

– **Efficiency**: Observe that

$$\sum_{i=1}^{n} \mu_i(u, \boldsymbol{x}) = \sum_{i=1}^{n} \sum_{j=1}^{\rho_i} \frac{\sigma_u^j(\boldsymbol{x}) - \sigma_u^{j-1}(\boldsymbol{x})}{n-j+1} = \sigma_u^n(\boldsymbol{x}) = \max_{i \in [n]} \sigma_u(x_i) = v\left([n]; \boldsymbol{x}\right).$$

– **Null player**: If $i$ is a null player, it follows that $v_u(\mathcal{C}; \boldsymbol{x}) = v_u(\mathcal{C} \cup \{i\}; \boldsymbol{x})$ for every coalition $\mathcal{C}$, and in particular, for $\mathcal{C} = \emptyset$. Therefore $v(\{i\}; \boldsymbol{x}) = v(\emptyset; \boldsymbol{x}) \triangleq 0$, hence $\sigma_u(x_i) = 0$. By definition of $\mu$, it holds that $\mu_i(u, \boldsymbol{x}) = 0$.
– **Symmetry**: $v(\{i\}; \boldsymbol{x}) = v(\{j\}; \boldsymbol{x})$ implies $\sigma_u(x_i) = \sigma_u(x_j)$, thus $\mu_i(u, \boldsymbol{x}) = \mu_j(u, \boldsymbol{x})$.
– **Linearity**: Note that $\mu$ is defined for a single user only. Therefore, we hereby extend it: for a distribution $f$ over $\{u_1, u_2\}$, define

$$\mu_i\left((\{u_1, u_2\}, f), \boldsymbol{x}\right) = f(u_1)\mu_i(u_1, \boldsymbol{x}) + f(u_2)\mu_i(u_2, \boldsymbol{x}).$$

Hence, linearity holds as well.

Since $\mu$ satisfies Shapley's axioms, $\mu_i\left(u, \boldsymbol{x}\right) = \phi_i\left(v_u(;\boldsymbol{x})\right)$, and $\mu_i\left((\mathcal{U}, f), \boldsymbol{x}\right)$ is the Shapley value of player $i$ in the cooperative game $V(;\boldsymbol{x})$. Moreover,

$$\mu_i\left((\mathcal{U}, f), \boldsymbol{x}\right) = \int_{\mathcal{U}} f(u)\mu_i(u, \boldsymbol{x})du = \pi_i(\boldsymbol{x}).$$

Thus the theorem is proved.

□

## 6 Discussion

We introduced Shapley facility location games, a framework incorporating probabilistic user behavior in facility location games. In this framework we considered choice selection among facilities motivated by the behavioral economics literature. Our results show that such probabilistic choice is "fair", and coincides with the Shapley value of a corresponding cooperative game. We proved that Shapley facility location games always possess pure Nash equilibria. We also crystallized the convergence rate in these games, and bounded their price of anarchy.

The reader may wonder whether the model can accommodate an asymmetric attraction function; that is, the case where the extent to which a user is attracted to a player depends not only on her chosen location, but also on her identity. Such asymmetry may result from power or influence a player possesses, which is a very natural assumption. Moreover, asymmetry can take the form of different sets of locations available to each player.

Consider a space $\mathcal{U}$ and a sequence of sets $\mathcal{L}_1, \mathcal{L}_2, \ldots, \mathcal{L}_n$, such that each player $i$ is limited to select a location in $\mathcal{L}_i$. For each player $i$, we define $\mathcal{S}_i : \mathcal{U} \times \mathcal{L}_i \to [0,1]$ to be the similarity function with respect to player $i$, where again we require $\mathcal{S}_i$ to be continuous or simple.

All the results obtained are carried on to the asymmetric extension with minor modifications. This is apart from the rate of the convergence to approximate Nash equilibria via best response dynamics, as games are no longer symmetric. In particular, a pure Nash equilibrium is still guaranteed to exist, the $PoA$ bound is still valid, and player payoffs correspond to Shapley values in the cooperative game.

An instance of such an asymmetric game, the *limited attraction model*, was recently discussed in [13,29]. In that model, the attraction of each player $i$ is limited to a ball of size $r_i$, and users outside her chosen ball will not be attracted to her. Thereupon, each user chooses, with equal probability, a player that attracts him. It can be verified that if $\mathcal{L}_i = \mathcal{U}$ and if the similarity function of player $i$ is

$$\forall u \in \mathcal{U}, l \in \mathcal{L}_i : \mathcal{S}_i(t,l) = \begin{cases} 1 & d(u,l) \leq r_i \\ 0 & Otherwise \end{cases},$$

the model obtained is exactly the model of [13]. In particular, it can be verified that player payoffs in [13] correspond to their Shapley value in the cooperative game introduced in the previous section.

Another interesting question is whether *every* Shapley facility location game possesses a unique pure Nash equilibrium, as it was the case in our illustrative example. Clearly, this is not the case. Taken to the extreme, consider a similarity function which is constant for every user and every location. It follows that every strategy profile is in equilibrium.

It is worth noticing that our work is distinguished from most previous work in facility location games, as our games are not zero-sum. Interestingly, we showed they are potential games [23,19], which allows us to connect to a main branch of research in the interplay between CS and game theory [20].

As for future work, we believe that putting data science tasks in the context of competition may be of interest. Since our model is general, tractable and efficient, it may serve as a benchmark for the study of strategic product selection in data science settings. Such settings include several Internet applications, e.g. where facilities and users are associated with document contents and queries, respectively, and the aim of the players (content authors) is to be the closest in their published content to as many queries as possible [4,15].

## Acknowledgments

This project has received funding from the European Research Council (ERC) under the European Union's Horizon 2020 research and innovation programme (grant agreement n° 740435).

# A Omitted Proofs

## A.1 Proof of Proposition 1

According to Theorem 1, every Shapley facility location game is a potential game. We shall use the potential function in order to describe the properties of the set of pure Nash equilibria of $\mathcal{G}$. Since $\mathcal{G}$ is symmetric, the potential function is permutation invariant, i.e. $\Phi(x_1, x_2) = \Phi(x_2, x_1)$.

Hence, w.l.o.g. we analyze the set of profiles $C = \{x_1, x_2 \in [0,1] : x_1 \leq x_2\}$. Observe that $C$ is a convex set.

Notice that $\Phi(\boldsymbol{x}) = \int_0^1 \left(\max\{\sigma_u(\boldsymbol{x})\} + \frac{1}{2}\min\{\sigma_u(\boldsymbol{x})\}\right) du$; hence $\Phi(\boldsymbol{x})$ can be computed directly using the area under the graph of the functions in Figure 2. It can be verified that this area is

$$\Phi(x_1, x_2) = x_2 + \frac{x_1}{2} - \frac{7}{8}(x_1^2 + x_2^2) + \frac{x_1 x_2}{4} + \frac{3}{4}.$$

Note that $\Phi(\boldsymbol{x})$ is strictly concave, hence attains one global maximum over $C$, which corresponds to a unique pure Nash equilibrium of $\mathcal{G}$. By differentiating $\Phi$ we find that maximum is obtained for $(x_1, x_2) = \left(\frac{3}{8}, \frac{5}{8}\right)$. □

### A.2 Proof of Theorem 3

For simplicity of notation, we use $\boldsymbol{x}$ to represent not only a strategy profile but also as the union of the corresponding sets of facilities, i.e. $\boldsymbol{x} = \bigcup_{i=1}^{t}\{x_n\}$. Next, we define the following set function:

$$v_u(\boldsymbol{x}) = \sum_{i=1}^{n} \mu_i(u, \boldsymbol{x}) = \max_{i \in [n]} \sigma_u(x_i).$$

In addition, we shall treat the social welfare as a set function as well:

$$V(\boldsymbol{x}) = \int_{\mathcal{U}} f(u) v_u(\boldsymbol{x}) du.$$

The proof of the Theorem relies on the following lemma:

**Lemma 4.** *For all $\boldsymbol{x} = (x_i, \boldsymbol{x}_{-i})$ it holds that:*

$$\pi_i(\boldsymbol{x}) \geq \frac{V(x_i)}{n} + \frac{n-1}{n}\left(V(\boldsymbol{x}) - V(\boldsymbol{x}_{-i})\right). \tag{8}$$

*Proof.* We distinguish between two complementary cases: if $v_u(x_i) > v_u(\boldsymbol{x}_{-i})$, then

$$\begin{aligned}
\mu_i(u, \boldsymbol{x}) &\geq \frac{1}{n} v_u(\boldsymbol{x}_{-i}) + v_u(x_i) - v_u(\boldsymbol{x}_{-i}) \\
&= v_u(x_i) - \frac{n-1}{n} v_u(\boldsymbol{x}_{-i}) \\
&= \frac{1}{n} v_u(x_i) + \frac{n-1}{n}\left(v_u(\boldsymbol{x}) - v_u(\boldsymbol{x}_{-i})\right),
\end{aligned}$$

where the last step follows from $v_u(x_i) = v_u(\boldsymbol{x})$ if $v_u(x_i) > v_u(\boldsymbol{x}_{-i})$. Alternatively, if $v_u(x_i) \leq v_u(\boldsymbol{x}_{-i})$, then $v_u(\boldsymbol{x}_{-i}) = v_u(\boldsymbol{x})$ and

$$\mu_i(u, \boldsymbol{x}) \geq \frac{1}{n} v_u(x_i) = \frac{1}{n} v_u(x_i) + \frac{n-1}{n}\left(v_u(\boldsymbol{x}) - v_u(\boldsymbol{x}_{-i})\right).$$

Therefore:

$$\begin{aligned}
\pi_i(\boldsymbol{x}) &= \int_{\mathcal{U}} f(u) \mu_i(u, \boldsymbol{x}) du \\
&\geq \int_{\mathcal{U}} f(u) \left(\frac{1}{n} v_u(x_i) + \frac{n-1}{n}\left(v_u(\boldsymbol{x}) - v_u(\boldsymbol{x}_{-i})\right)\right) \\
&= \frac{V(x_i)}{n} + \frac{n-1}{n}\left(V(\boldsymbol{x}) - V(\boldsymbol{x}_{-i})\right).
\end{aligned}$$

□

One more necessary notion is submodularity.

**Definition 3 (Submodular function).** *We say that* $f : 2^{\Omega} \to \mathbb{R}$ *is submodular if for every* $A \subseteq B \subseteq \Omega$ *and every* $\omega \in \Omega \setminus B$ *it holds that -*

$$f(A \cup \{\omega\}) - f(A) \geq f(B \cup \{\omega\}) - f(B).$$

**Lemma 5.** *For every* $u \in \mathcal{U}$ *,* $v_u$ *is submodular.*

*Proof.* Fix arbitrary $u \in \mathcal{U}$. We need to show that

$$v_u\left(\boldsymbol{x}_{-i} \cup \{\omega\}\right) - v_u\left(\boldsymbol{x}_{-i}\right) \geq v_u\left(\boldsymbol{x} \cup \{\omega\}\right) - v_u\left(\boldsymbol{x}\right). \tag{9}$$

Observe that $v_u(\boldsymbol{x})$ is monotonically non-decreasing, thus $v_u(\boldsymbol{x}) \geq v_u(\boldsymbol{x}_{-i})$. If $v_u\left(\{\omega\}\right) \leq v_u(\boldsymbol{x}_{-i})$, then both sides of Equation (9) are zero. Else, if $v_u(\boldsymbol{x}_{-i}) < v_u\left(\{\omega\}\right) \leq v_u\left(\boldsymbol{x}\right)$, the right-hand side of Equation (9) is zero, while the left-hand side is positive. Ultimately, if $v_u\left(\{\omega\}\right) > v_u\left(\boldsymbol{x}\right)$,

$$v_u\left(\boldsymbol{x}_{-i} \cup \{\omega\}\right) - v_u\left(\boldsymbol{x}_{-i}\right) = v_u\left(\boldsymbol{x} \cup \{\omega\}\right) - v_u\left(\boldsymbol{x}_{-i}\right) \geq v_u\left(\boldsymbol{x} \cup \{\omega\}\right) - v_u\left(\boldsymbol{x}\right).$$

□

**Corollary 2.** *The social welfare function* $V$ *is submodular.*

From here on, we follow the technique presented in [31] of *valid utility systems*, later crystallized in [24]. Denote the optimal solution $\boldsymbol{x}^*$ and equilibrium profile $\boldsymbol{x}$. Due to Lemma 4 we have:

$$\pi_i(x_i^*, \boldsymbol{x}_{-i}) \geq \frac{V(x_i^*)}{n} + \frac{n-1}{n}\left(V(x_i^*, \boldsymbol{x}_{-i}) - V(\boldsymbol{x}_{-i})\right). \tag{10}$$

By summing Equation (10) over all players:

$$\sum_{i=1}^{n} \pi_i(x_i^*, \boldsymbol{x}_{-i}) \geq \sum_{i=1}^{n} \frac{V(x_i^*)}{n} + \frac{n-1}{n} \sum_{i=1}^{n}\left(V(x_i^*, \boldsymbol{x}_{-i}) - V(\boldsymbol{x}_{-i})\right). \tag{11}$$

Due to submodularity of $V$, for every $i$ it holds that

$$V(x_i^*, \boldsymbol{x}_{-i}) - V(\boldsymbol{x}_{-i}) \geq V(x_1^*, \ldots, x_{i-1}^*, x_i^*, \boldsymbol{x}) - V(x_1^*, \ldots, x_{i-1}^*, \boldsymbol{x}).$$

Thus

$$\begin{aligned}
\sum_{i=1}^{n}\left(V(x_i^*, \boldsymbol{x}_{-i}) - V(\boldsymbol{x}_{-i})\right) &\geq \sum_{i=1}^{n}\big(V(x_1^*, \ldots, x_{i-1}^*, x_i^*, \boldsymbol{x}) - V(x_1^*, \ldots, x_{i-1}^*, \boldsymbol{x})\big) \\
&= V(\boldsymbol{x}^*, \boldsymbol{x}) - V(\boldsymbol{x}) \\
&\geq V(\boldsymbol{x}^*) - V(\boldsymbol{x}). \qquad (12)
\end{aligned}$$

Since $\boldsymbol{x}$ is in equilibrium, it follows that $\pi_i(\boldsymbol{x}) \geq \pi_i(x_i^*, \boldsymbol{x}_{-i})$. Combining Equations (11) and (12), we have

$$\begin{aligned}
V(\boldsymbol{x}) &= \sum_{i=1}^{k} \pi_i(\boldsymbol{x}) \geq \sum_{i=1}^{k} \pi_i(x_i^*, \boldsymbol{x}_{-i}) \\
&\geq \sum_{i=1}^{n} \frac{V(x_i^*)}{n} + \frac{n-1}{n} \left(V(\boldsymbol{x}^*) - V(\boldsymbol{x})\right) \\
&\geq \frac{V(\boldsymbol{x}^*)}{n} + \frac{n-1}{n} \left(V(\boldsymbol{x}^*) - V(\boldsymbol{x})\right) \\
&= V(\boldsymbol{x}^*) - \frac{n-1}{n} V(\boldsymbol{x}).
\end{aligned}$$

Ultimately:

$$PoA \triangleq \frac{V(\boldsymbol{x}^*)}{V(\boldsymbol{x})} \leq \frac{2n-1}{n}.$$